\documentclass[12pt]{iopart}
\usepackage{psfrag}

\usepackage{xspace}
\usepackage{graphics}
\usepackage{graphicx}
\usepackage{iopams}

\usepackage{amsfonts}
\usepackage{graphics}
\usepackage{psfrag}
\usepackage{hyperref}
\usepackage{graphicx}
\usepackage{epsfig}
\usepackage[latin1]{inputenc}
\usepackage{color}
\usepackage{rotating}

\newcommand{\gcomment}[1]{}

\begin{document}
\title[How target waves emerge in population dynamics]
{How target waves emerge in population dynamics with cyclical
interactions}
\author{Luo-Luo Jiang$^1$, Tao Zhou$^1$$^,$$^2$, Xin Huang$^3$, Bing-Hong Wang$^1$}

\address{$^1$\ Department of Modern Physics, University of Science and
Technology of China, Hefei 230026, PR China}
\address{$^2$ Department of Physics, University of Fribourg, Chemin du Muse
3, CH-1700 Fribourg, Switzerland}
\address{$^3$\ Department of Physics, University of Science and
Technology of China, Hefei 230026, PR China}

\date{\today}

\ead{bhwang@ustc.edu.cn}

\begin{abstract}
Based on a multi-agent model, we investigate how target waves
emerge from a population dynamics with cyclical interactions among
three species. We show that the periodically injecting source in a
small central area can generate target waves in a two-dimensional
lattice system. By detecting the temporal period of species'
concentration at the central area, three modes of target waves can
be distinguished. Those different modes result from the
competition between local and global oscillations induced by
cyclical interactions: $Mode~A$ corresponds to a synchronization
of local and global oscillations, $Mode~B$ results from an
intermittent synchronization, and $Mode~C$ corresponds to the case
when the frequency of the local oscillation is much higher than
that of the global oscillation. This work provides insights into
pattern formation in biologic and ecologic systems that are
totally different from the extensively studied diffusion systems
driven by chemical reactions.
\end{abstract}


\maketitle

\section{Introduction}
Spatially distributed excitable systems are widely investigated
owning to their biological significance of long-rang signal
transmission through self-sustained waves
\cite{inti1,inti2,inti3,inti4,inti5,inti6}. Pattern formation of
wave propagation in excitable systems have been well studied
\cite{inti7}. Especially, target waves induced by noise have been
found in the recurrent single species' population dynamics
\cite{inti6}. However, the formation mechanism of target waves in
populations dynamics driven by multi-species' competitive
interactions are not very clear.

Competitive interactions in natural and social systems consisted
of many elements (e.g., various biological species, political
parties, businesses, coupled reaction chemical components,
bacterial production bacteria, etc.) play an important role in
evolutionary processes and pattern formations
\cite{inta1,inta2,inta3,inta4,inta5,inta6}, and lead to the
emergence of spatial patterns. In particular, target waves are
commonly observed in those systems
\cite{intf1,intf2,intf3,intf4,intf5}. Previous works reveal that
patterns could emerge in reaction-diffusion systems if one of the
substances diffuses much faster than others
\cite{inte1,inte2,inte3}. For example, the bromous acid diffuses
much faster than ferrion in the Belousev-Zhabotinsky (BZ) reaction
and the cyclic AMP diffuses much faster than membrane receptor in
the dictyostelium discoideum. However, many pattern formations of
mobile population in ecosystems, such as migrating animals and
running bacteria, can not be explained by the reaction-diffusion
mechanism, since the diffusion speed induced by individual
mobility is the same for all substances \cite{intd1,intdd1}.
Furthermore, partial differential equations (PDEs) have been
proposed to describe the evolution of pattern formation in
reaction-diffusion systems, such as the Oregonator model for BZ
reaction \cite{intc2}. Based on those PDEs, one can analyze the
stability of patterns by using the mean field theory, but the
reliable detailed information is quite limited. Therefore, it is
necessary to use an agent-based models to describe the pattern
formation of mobile population \cite{inta4}.

The cyclic predator-prey model provides a terse description of
competition among population of different species. Experimental
study \cite{intd1} has revealed that the mechanism of
rack-paper-scissors game can promote the biodiversity of three
strains of $\emph{Escherichia coli}$. Very recently, Reichenbach
\emph{et al} proposed a rack-paper-scissors game of mobile
populations \cite{inta1}, where the individual mobility displays a
critical effect on species diversity. When mobility is below a
certain value, all species coexist and form spiral waves. In
contrast, above this threshold biodiversity is jeopardized.
Indeed, previous investigations show that the cyclical competition
mechanism and low mobility of individuals maintain the biodivesity
in ecosystems \cite{inta1,intc1,intd1,intd2,intd3}. In this paper,
we investigate the pattern formation based on a cyclic
predator-prey model with mobile individuals. Target waves are
observed from the recurrent dynamics driven by three species'
competitive interactions. As a result of global oscillation, a
transition from disordered state to an ordered spacial structure
occurs with the increasing of injection period in the vortex of
target waves. By detecting the temporal period of species'
concentration at the central area, three modes of target waves can
be distinguished. Those different modes result from the
competition between local and global oscillations induced by
cyclical interactions: $Mode~A$ corresponds to a synchronization
of local and global oscillations, $Mode~B$ results from an
intermittent synchronization, and $Mode~C$ corresponds to the case
when the frequency of the local oscillation is much higher than
that of the global oscillation. Though the present model only
concerns the microscopic interactions among individuals, target
waves emerge in a macroscopic level. Furthermore, our work
provides a feasible method for pattern control and pattern
selection.

\section{Stochastic cyclic predator-prey model}

Based on the previous works of Reichenbach \emph{et al}
\cite{inta1, intc1}, we introduce a cyclic predator-prey model as
follow: Mobile individuals of three species (marked by 1, 2, 3)
locate in the nodes of a two-dimensional regular lattice with no
flux boundary. Here a node stands for an $1\times1$ square. Each
node can be occupied by at most one individual of a species or a
vacancy (denoted by $V$) representing resource. There are three
kinds of interactions, namely exchange, predation, and
reproduction, which only occur between neighboring nodes.
$\emph{Exchange.}$--- An individual could exchange positions with
one of its neighbors at a rate $\alpha$ due to its mobility.
$\emph{Predation.}$--- $1$ beats $2$ and $2$ becomes a vacancy at
a predation intensity $\beta$, in the same way, $2$ beats $3$, and
$3$ beats $1$. $\emph{Reproduction.}$--- An individual can
reproduce an offspring to a neighboring $V$ node at a rate
$\gamma$.

Unlike the deterministic approach which regards the time evolution
as a continuous process, here, the applied stochastic approach
regards the time evolution as a kind of random-walk process. In
this model, reactions occur in a random manner: exchange happens
with probability $\alpha /(\alpha +\beta+\gamma)$, predation with
probability $\beta /(\alpha +\beta+\gamma)$, and reproduction with
probability $\gamma /(\alpha +\beta+\gamma)$. We use a standard
algorithm for stochastic simulation, which was developed by
Gillespie \cite{stoa1,stoa2}.

\section{Results}
\begin{figure}
\begin{center}
\scalebox{1.0}[1.0]{\includegraphics{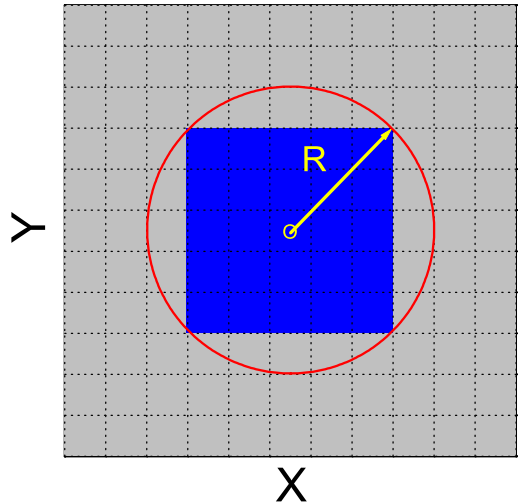}}
\end{center}
\caption{Illustration of the injected area in an $11\times11$
lattice. Here, $R$ is set as $3.5$,  and the lattice completely
inside the circle, marked by the blue color, is the injected area.
In the simulation of this paper, the injected radius $R$ is fixed
as $10.5$. \label{injection}}
\end{figure}

Inspired by the experiments of growth of $\emph{Escherichia coli}$
on a petri dish \cite{intf2}, we apply the dynamical model of
three populations' cyclical interactions \cite{inta1}. In an
$L\times L$ lattice, each node presents a space occupied by a
vacancy or an individual of a species, and at each time step a
randomly chosen individual interacts with one of its four nearest
neighbors, which is also determined randomly according to the
corresponding predation intensity. The evolving process is
implemented by using the Gillespie algorithm \cite{stoa1,stoa2} in
which one Monte Carlo (MC) time is defined as the time period
during which all the individuals have been chosen once on average.
Hereinafter, the resolution of time is defined as one MC time.
Initially, the lattice is wholly occupied by vacancies, and then
populations of a species are injected periodically in a central
area, namely the \emph{injected area}. An illustration of the
injected area is shown in Fig.~\ref{injection}. In this paper, the
injected radius, $R$, is set as $10.5$ and the lattice size, $L$,
is $1024$. The injected species revolve in the order $1$, $2$,
$3$, $1$, $2$, $3$, ... For example, populations of $1$ are
injected at $t=0$, then $2$ at $t=T_{0}$, $3$ at $t=2T_{0}$, and 1
at $t=3T_{0}$, etc. In this case, the period of injecting is
$T_{in}=3T_{0}$, where $T_{0}$ is the interval time between
injections of two species.
\begin{figure}
\begin{center}
\scalebox{0.6}[0.6]{\includegraphics{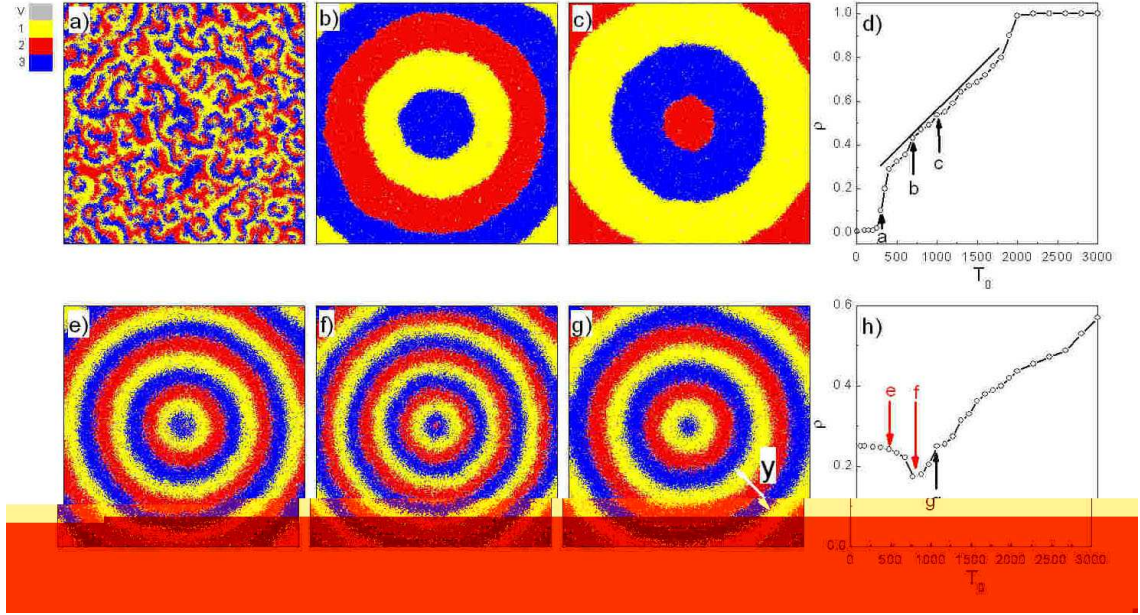}}
\end{center}
\caption{The panels (a), (b), (c), (e), (f) and (g) display
different pattern formations generated in the present model. The
panels (d) and (h) report $\rho$ as a function of $T_{0}$. (a),
(b), (c), (d) for $M=10^{-5}$ and (e), (f), (g), (h) for
$M=10^{-4}$. (a) and (e) for $T_{0}=300$, (b) and (f) for
$T_{0}=500$, (c) and (g) for $T_{0}=1000$. The cases presented in
(a), (b), and (c) are marked by by $a$, $b$, and $c$ in (d), and
the cases presented in (e), (f), and (g) are marked by $e$, $f$,
and $g$ in $(h)$. In (d), the relation $\rho({T_{0}})$ can be
approximately linearly fitted in the interval $T_{0} \in (500,
1800)$. In (g), $y$ is denotes the spatial distance between
neighboring wave fronts for the same species. \label{pattern}}
\end{figure}

\begin{figure}
\begin{center}
\scalebox{0.7}[0.8]{\includegraphics{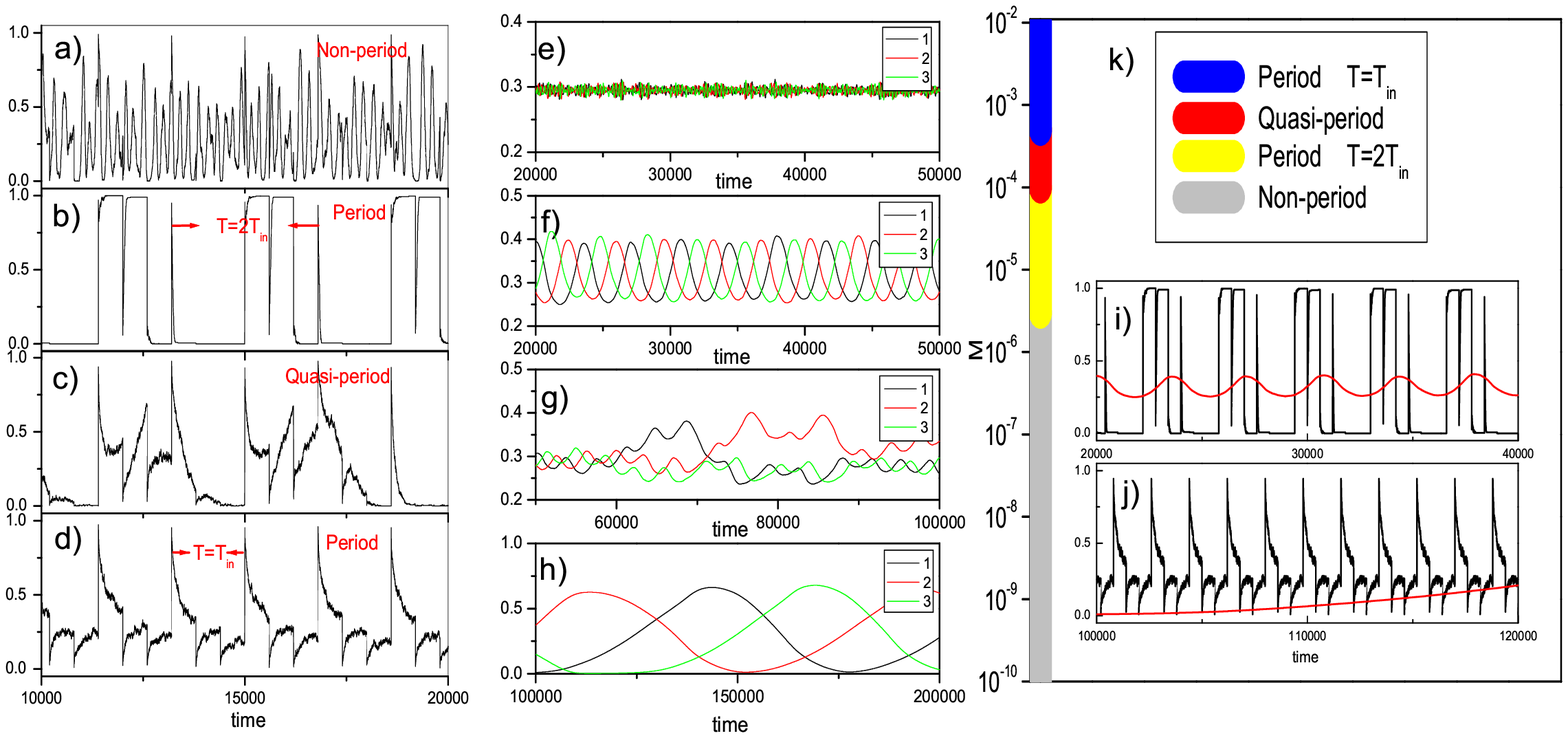}}
\end{center}
\caption{The panels (a)-(d) report the proportion of species $1$
in the injected area for different mobilities, while (e)-(h)
display proportions of three species in the whole system. Panels
(e)-(h) are one-to-one corresponding to panels (a)-(d). $T_{0}$ is
fixed as 600, $M=10^{-6}$ for (a) and (c), $M=10^{-5}$ for (b) and
(f), $M=10^{-4}$ for (c) and (g), $M=10^{-3}$ for (d) and (h). The
panel (k) shows different system states from $M=10^{-10}$ to
$M=10^{-2}$, given $T_{0}=600$. The insets (i) and (j) display the
comparison of species $1$'s proportions in the injected area
(black) and the whole system (red),(i) for $M=10^{-5}$
(corresponding to (b) and (f)) and (j) for $M=10^{-3}$
(corresponding to (d) and (h)). \label{proportion}}
\end{figure}

Mobility is an important character for most population dynamics
such as bacteria swimming and tumbling. The mobility in the
present model corresponds to the action that an individual moves
to a neighboring empty node ($V$ node), which is defined as:

\begin{eqnarray}
M=2\alpha /N,
\end{eqnarray}
where $N=L^2$ is the number of nodes in the system. In this paper,
$L=1024$, $\beta=1$ and $\gamma=1$ are fixed.

With no-flux boundary conditions, extensive computer simulations
are performed and the system displays abundance phenomena of
pattern formation with various injection periods. In order to
describe the spatial structure of pattern formation, we define
$\rho=N_{0}/N$, where $N_{0}$ is the number of individuals in the
largest \emph{adjacent population}. Here, an adjacent population
is a set of individuals of one species where any two individuals
can be connected through a sequence of nearest neighboring
relations within this set. Apparently, patterns break when $\rho$
approaches to $0$, while one species occupies the whole system
when $\rho$ reaches $1$. As shown in Fig.~\ref{pattern}(d), given
$M=10^{-5}$, with the increasing of $T_{0}$, the system undergoes
a transition from a state of coexistence of three species to a
state occupied by one species. The pattern formations
corresponding to the points $a$, $b$, and $c$ in
Fig.~\ref{pattern}(d) are shown in
Fig.~\ref{pattern}(a),~\ref{pattern}(b), and~\ref{pattern}(c)
respectively. When $T_{0}$ is small ($T_{0}\leq 300$) the three
species form many small spiral waves which can be considered as a
disordered state in global level. Fig.~\ref{pattern}(b) and
Fig.~\ref{pattern}(c) show target waves emerging for $T_{0}=500$
and $T_{0}=1000$. The system becomes more and more ordered,
reaches a consensus state when $T_{0}\geq 2000$. In contrast, for
$M=10^{-4}$, Fig.~\ref{pattern}(h) shows an unexpected decline
with the increasing of $T_{0}$, and the pattern formations
corresponding to the points $e$, $f$, $g$, are shown in
Fig.~\ref{pattern}(e),~\ref{pattern}(f), and~\ref{pattern}(g),
respectively.

It is worth mentioning that all patterns in our results are robust
after the systems have reached steady states. It seems that target
waves emerge because of the increasing of $T_{in}$. A question is
naturally taken into consideration: How $T_{in}$ affects the
pattern formation. Furthermore, one may ask why $\rho$ is
monotonously increasing with the increasing of $T_{0}$ in
Fig.~\ref{pattern}(d), while non-monotonously varying in
Fig.~\ref{pattern}(h). We calculate the proportion of species $1$
in the injected area. With $T_{0}=600$, this proportion does not
display periodicity for $M=10^{-6}$ (see
Fig.~\ref{proportion}(a)), while varies periodically for
$M=10^{-5}$ (see Fig.~\ref{proportion}(b)) as well as for
$M=10^{-3}$ (see Fig.~\ref{proportion}(d)), and shows
quasi-periodical behavior for $M=10^{-4}$ (see
Fig.~\ref{proportion}(c)). The period is $T=2T_{in}=3600$ for
$M=10^{-5}$, and $T=T_{in}=1800$ for $M=10^{-3}$.
Fig.~\ref{proportion}(e),~\ref{proportion}(f),~\ref{proportion}(g),
and~\ref{proportion}(h) show (the varying of) three species'
proportions in the whole system one-to-one corresponding to the
evolving processes of
Fig.~\ref{proportion}(a),~\ref{proportion}(b),~\ref{proportion}(c),
and~\ref{proportion}(d). Clearly,
Fig.~\ref{proportion}(a)-\ref{proportion}(d) show the local
oscillations in the injected area, while
Fig.~\ref{proportion}(e)-\ref{proportion}(h) display the global
oscillations. Fig.~\ref{proportion}(i) and
Fig.~\ref{proportion}(j), corresponding to the cases shown
in~\ref{proportion}(b) and~\ref{proportion}(d), compared the local
and global oscillations of species $1$. In
Fig.~\ref{proportion}(i), the local and global oscillations have
the same period, while in Fig.~\ref{proportion}(j), the frequency
of the local oscillation is much higher than that of the global
oscillation. We say, in the case shown in
Fig.~\ref{proportion}(b), ~\ref{proportion}(f) and
\ref{proportion}(i), the local and global oscillations are
synchronous at $T=2T_{in}$, while in the case shown in
Fig.~\ref{proportion}(d), ~\ref{proportion}(h) and
\ref{proportion}(j), they are non-synchronous. In between, the
system displays quasi-periodical behavior, and the local and
global oscillations are intermittent synchronous (see, for
example, Fig.~\ref{proportion}(c) and~\ref{proportion}(g)).
According to the above discussion, a qualitative classification of
system states versus mobility is shown in
Fig.~\ref{proportion}(k), including non-period, period
$T=2T_{in}$, quasi-period, and period $T=T_{in}$.

\begin{figure}
\begin{center}
\includegraphics*[scale=0.5]{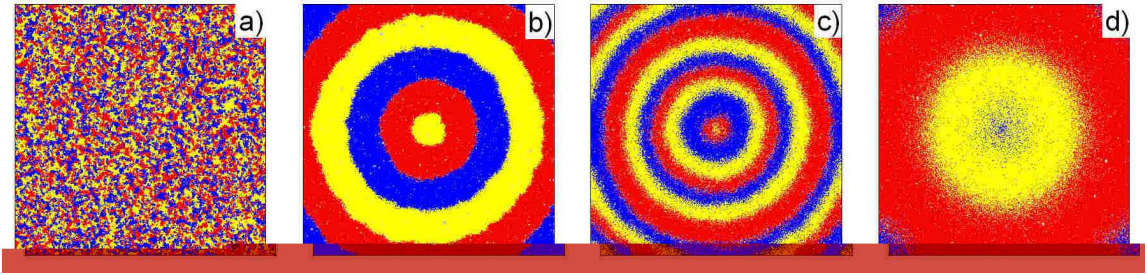}
\end{center}
\caption{The panels (a)-(d) show pattern formations corresponding
to the cases shown in Fig.~\ref{proportion}(a), (b), (c), and (d)
respectively. \label{target}}
\end{figure}

As shown in Fig.~\ref{target}, there are four qualitatively
different states, and the patterns in Fig.~\ref{target}(b),
Fig.~\ref{target}(c) and Fig.~\ref{target}(d) are named as
$Mode~A$, $Mode~B$ and $Mode~C$. When a species is injected,
individuals of this species will interact with other species
around. $(i)$ For extremely small value of $M$ ($<3.0\times
10^{-6}$), individuals out of the injected area distribute almost
homogeneously due to a relatively high rate of predation and
reproduction. The injected individuals, surrounded by other two
species, are segmented into fragments. Therefore, as shown in
Fig.~\ref{target}(a) the local oscillations have no period, and no
global ordered pattern formats. $(ii)$ For small value of $M$
($<10^{-4}$), individuals out of the injected area are
inhomogeneously distributed due to a relatively low rate of
predation and reproduction. The injected individuals are
surrounded by only the prey species, and each species takes up the
injected area for $2T_{in}/3$, thus the period of local
oscillations is 2$T_{in}$. The boundary between the two species
near the injected area moves from the center to the whole system,
inducing a global oscillation with period 2$T_{in}$ too. This case
is shown in Fig.~\ref{target}(b) and referred to as $Mode~A$.
$(iii)$ For medium value of $M$ ($10^{-4}< M < 5.0\times
10^{-4}$), although individuals out of the injected area are
inhomogeneously distributed due to a relatively low rate of
predation and reproduction, the injected individuals are
surrounded by all three species. No species takes up the whole
injected area and the period of local oscillations is 2$T_{in}$
sometimes (see Fig.~\ref{proportion}(c)). In other words, local
oscillations are confined in the injected area sometimes.
Therefore, the local oscillations is quasi-periodic and
intermittently synchronized with the global oscillations. This
case is shown in Fig.~\ref{target}(c) and referred to as $Mode~B$.
$(iv)$ When $M$ is sufficiently large ($>5.0\times 10^{-4}$), the
relative rate of predation and reproduction is very low. Three
species coexist near the boundary of the injected area. Because
the number of injected individuals is quickly depressed, no
species dominates the whole injected area (see
Fig.~\ref{proportion}(d)). Local oscillations are confined in the
injected area, and individuals of a species will aggregate near
the boundary of the injected area, and then move from the center
to the whole system. Since the aggregation and propagation takes a
long time, the period of global oscillations is much longer than
that of local oscillations. Target waves formatting in this manner
are called $Mode~C$.

The wavelength is defined as $\lambda=y/L$, where $y$ is denotes
the spatial distance between neighboring wave fronts with the same
species in the lattice as shown in Fig.~\ref{pattern}(g).
Comparing Fig.~\ref{target}(b), Fig.~\ref{target}(c), and
Fig.~\ref{target}(d), one can find that wavelengths of target
waves in both $Mode~A$ and $Mode~C$ are longer than that in
$Mode~B$, and wavelengths of target waves in $Mode~B$ are
inequable. It can explain why Fig.~\ref{pattern}(h) have a trop.
The reason is target waves belong to $Mode~B$ for $T_{0}>300$ and
$T_{0}\leq 800$ at $M=10^{-4}$, while others belong to $Mode~A$ or
$Mode~C$, and wavelengths of $Mode~B$ are smaller than that of
$Mode~A$ and $Mode~C$. Moreover, the target waves become unstable
when $T_{0}$ goes across the critical point distinguishing two
different modes, and may evolve into spiral waves which is also
observed in a system of a single species, \emph{Dictyostelium}
system \cite{inti6,con1a}.

\begin{figure}
\begin{center}
\includegraphics*[scale=0.8]{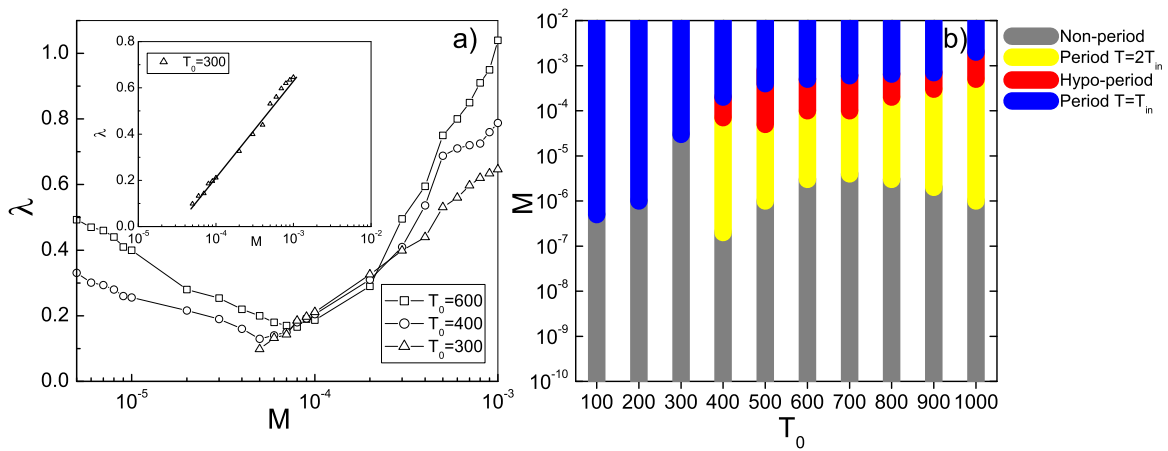}
\end{center}
\caption{(a) Wavelength of target waves as a function of $M$ for
different values of injection period $T_{0}$. The inset shows
$\lambda \sim logM^{1/2}$. (b) Different system states in
$M-T_{0}$ space.\label{wavelength}}
\end{figure}

As shown in Fig.~\ref{wavelength}(a), wavelengths of target waves
display non-monotonic behavior with the mobility of individuals
for $T_{0}=600$ and $T_{0}=400$, while increase with $M$ in terms
of $\lambda\sim \log M^{1/2}$ for $T_{0}$=300. This phenomena are
different from the cases of spiral waves \cite{inta1}, in which
$\lambda \sim M^{1/2}$. The non-monotonic increasing with the
mobility $T_{0}=400$ and $T_{0}=600$ is caused by the target waves
of $Mode~B$ (see red blocks in Fig.~\ref{wavelength}(b)).

\section{Summary}

Cyclical competition, an important principle of ecology and society,
leads to complex phenomena of maintenance of ecological
bio-diversity and emergence of cooperation in society, where spatial
distribution of populations plays a critical role
\cite{conb1,conb2,conb3,conb4}. It has been found that
multi-species' competition induces complex spatiotemporal structures
in ecological systems. Especially, the 3-species evolving game,
namely rock-paper-scissors game or cyclic predator-prey model,
reveals spiral wave structures \cite{conb7,conb8}. However, target
waves are rarely reported. In the current model, target waves emerge
under a suitable initial condition, which is in accordance with the
observations in real bacteria systems \cite{inti6,con1a}. Although
target waves were also observed in many excitable media, in our
model, individuals of three species move at the same rate, which
distinguishes our model from the diffusion systems driven by
chemical reactions where different substances have different moving
speeds. Indeed, our results suggest that target waves can be driven
by the competition of local and global oscillations. This newly
reported mechanism could provide insights into target waves
formation in biologic and ecologic systems.

In addition, our results can be applied in pattern control.
Actually, the system makes stochastic resonance to the periodical
injection fluid in $Mode~A$, which can be used to control the global
behaviors since the injection period is controllable. Recently, the
periodic injection method is also used in complex Ginzburg-Landau
system for the purpose of spatiotemporal chaos control
\cite{con1,con2}. The periodical injection for pattern control has
great significance in potential applications of toothful medication
transportation.

\ack{ The authors would like to thank Dan Peng, Tian-Shuo Hu,
Xiao-Pu Han, and Shi-Min Cai for their assistances in preparing
this manuscript. This work is supported by the National Basic
Research Program of China (973 Program No. 2006CB705500), the
National Natural Science Foundation of China (Grant Nos. 10635040,
10532060 and 10472116), and by the Specialized Research Fund for
the Doctoral Program of Higher Education of China. }

\section*{References}

\bibliographystyle{unsrt}


\end{document}